\documentclass{article}
\usepackage{spconf,amsmath,graphicx,hyperref}
\usepackage{tabularx}   
\usepackage{booktabs}
\usepackage{graphicx} 
\usepackage{multirow}
\usepackage{amssymb}   

\title{The ICASSP 2026 HumDial Challenge: Benchmarking Human-like Spoken Dialogue Systems in the LLM Era}

\name{
    \begin{tabular}{@{}c@{}}
    \itshape Zhixian Zhao$^{1*}$, Shuiyuan Wang$^{1*}$, Guojian Li$^{1*}$, Hongfei Xue$^{1*}$, Chengyou Wang$^{1*}$, \\ \itshape
    Shuai Wang$^{2}$, Longshuai Xiao$^{3}$, Zihan Zhang$^{3}$, Hui Bu$^{4}$, Xin Xu$^{4}$, \\ \itshape
    Xinsheng Wang$^{5}$, Hexin Liu$^{6}$, Eng Siong Chng$^{6}$, Hung-yi Lee$^{7}$, Lei Xie$^{1\dagger}$
    \thanks{*Equal contribution. $^{\dagger}$Corresponding author (lxie@nwpu.edu.cn).}
    \end{tabular}
}

\address{
    $^{1}$ASLP@NPU  $^{2}$NJU  $^{3}$Huawei Technologies  $^{4}$AISHELL  $^{5}$Soul AI lab  $^{6}$NTU, SG  $^{7}$NTU
}

\begin{document}
\ninept
\maketitle
\begin{abstract}
Driven by the rapid advancement of Large Language Models (LLMs), particularly Audio-LLMs and Omni-models, spoken dialogue systems have evolved significantly, progressively narrowing the gap between human-machine and human-human interactions. Achieving truly ``human-like'' communication necessitates a dual capability: \textit{emotional intelligence} to perceive and resonate with users' emotional states, and \textit{robust interaction} mechanisms to navigate the dynamic, natural flow of conversation, such as real-time turn-taking. Therefore, we launched the first Human-like Spoken Dialogue Systems Challenge (HumDial) \footnote{For more information and detailed results, visit: \url{https://aslp-lab.github.io/HumDial-Challenge/}} at ICASSP 2026 to benchmark these dual capabilities. Anchored by a sizable dataset derived from authentic human conversations, this initiative establishes a fair evaluation platform across two tracks: (1) Emotional Intelligence, targeting long-term emotion understanding and empathetic generation; and (2) Full-Duplex Interaction, systematically evaluating real-time decision-making under `` listening-while-speaking'' conditions. This paper summarizes the dataset, track configurations, and the final results.
\end{abstract}
\vspace{-9pt}
\section{Introduction}
\vspace{-7pt}
\label{sec:intro}

The rapid evolution of Large Language Models (LLMs) has driven a paradigm shift in spoken dialogue systems, transitioning from cascaded pipelines to unified Audio-LLM architectures. Proprietary models, exemplified by GPT-4o Realtime~\cite{GPT-4o}, have set a new benchmark for authentic interaction, demonstrating that native omni-modal modeling can deliver superior, low-latency speech-to-speech experiences. Following this trend, the open-source community has swiftly contributed competitive models, such as Qwen2.5-Omni~\cite{Qwen2.5}, GLM-4-Voice~\cite{GLM-4-Voice}, and many others~\cite{Kimi, vita, Step-Audio}. These advancements collectively demonstrate that modern dialogue systems can achieve impressive interaction quality, offering users a seamless experience that closely mimics the responsiveness of natural conversation.

Despite these significant strides, a fundamental question remains: \textit{How far are state-of-the-art human–machine dialogue systems from achieving human-level conversational naturalness?} While current models excel in task-completion, measuring their ability to replicate the subtle nuances of human communication—specifically deep emotional resonance and the implicit logic inherent in complex turn-taking—requires a standardized evaluation ground. To address this question and provide a fair benchmarking platform, we initiated the first Human-like Spoken Dialogue Systems Challenge (HumDial) at ICASSP 2026. This challenge is designed to assess two tracks of human-like interaction rigorously: (I) \textit{Emotional Intelligence}, focusing on multi-turn emotional trajectory tracking, causal reasoning, and empathetic response generation; and (II) \textit{Full-Duplex Interaction}, evaluating the system's ability to handle interruptions and maintain the natural flow of conversation via concurrent listening and generation. This paper summarizes the challenge outputs, including the dataset construction, track configurations, and the final results.

\vspace{-11pt}
\section{Related Work}
\vspace{-7pt}
With the rise of Audio-LLMs, emotion evaluation is shifting from simple recognition (e.g., EmoBench-m~\cite{EmoBench-m}) to deep emotional interaction. While common datasets~\cite{IEMOCAP, meld}, along with recent challenges~\cite{mer2024,meiju}, have enriched assessment dimensions, they largely adopt a ``static classification'' paradigm insufficient for dynamic emotional understanding and generation. Although ContextDialog~\cite{ContextDialog} and Multi-Bench~\cite{MULTI-Bench} introduce context, they often exhibit a ``pseudo-multi-turn'' nature—relying on concatenated single-turn dialogues or synthetic speech. This approach disrupts the natural flow of emotion, making it challenging to assess consistency across long-term interactions.
In real-time interaction, Full-Duplex-Bench~\cite{Full-Duplex-Bench} established a taxonomy of interruptions, whereas MTalk-Bench~\cite{MTalk-Bench} introduced metrics for paralinguistic cues and ambient noise. However, these benchmarks rely on ``synthetic mixing''—the artificial overlap of audio tracks. This mechanical approach fails to capture natural cognitive synchronization, such as hesitations or cooperative barges-in. Furthermore, existing work often overlooks rejection capability—the ability to remain silent amid background noise. To address these limitations, HumDial leverages real-world recorded dialogues to establish a unified standard for multi-turn emotional evolution and reasoning, as well as robust full-duplex interaction.

\begin{table*}[t]
\centering
\footnotesize
\caption{Statistics of the dataset across two tracks. * Task 3 consists of utterances sampled from Task 1 \& 2.}
\label{tab:dataset_stats}
\begin{tabular}{lccccccc} 
\toprule 
\multirow{2}{*}{\textbf{Split}} & \multicolumn{4}{c}{\textbf{Track I: Emotional Intelligence}} & \multicolumn{3}{c}{\textbf{Track II: Full-Duplex Interaction}} \\
\cmidrule(lr){2-5} \cmidrule(lr){6-8} 
 & \textbf{Task 1 \#Dia} & \textbf{Task 2 \#Dia} & \textbf{Task 3 \#Utt *} & \multirow{2}{*}{\textbf{Total \#Utt}} & \textbf{Interruption} & \textbf{Rejection} & \multirow{2}{*}{\textbf{Total \#Utt}} \\
 & (3/4/5 turns) & (3/4/5 turns) & \textit{(Sampled)} &  & (FQ/ND/RR/TS/ST) & (URB/PH/TPS/SDO) & \\ 
\midrule 
Train & 1600/1600/1600 & 1600/1600/1600 & -- & 38,400 & 1507/1211/1213/1213/1212 & 1211/1211/1210/0 & 9,418 \\
Dev   & 100/100/100    & 100/100/100    & 100/100/100 & 2,400  & 200/200/200/200/200     & 200/200/200/200 & 1,800 \\
Test  & 100/100/100    & 94/95/94    & 100/100/94 & 2,332  & 600/600/600/600/600     & 600/600/600/200 & 5,000 \\
\bottomrule 
\end{tabular}
\vspace{-13pt}
\end{table*}

\vspace{-13pt}
\begin{table*}[h]
\centering
\caption{Results of Track I - Emotional Intelligence}
\label{tab:track1_results}
\resizebox{\linewidth}{!}{
\begin{tabular}{l c c c c c c c}
\toprule
\textbf{Team} & \textbf{Task1 (D1/D2/D3)} & \textbf{Task1\_Avg} & \textbf{Task2 (D1/D2/D3)} & \textbf{Task2\_Avg} & \textbf{Task3 (D1/D2/D3)} & \textbf{Final Score} & \textbf{Rank} \\
\midrule
TeleAI             & 4.94/4.99/4.99 & 4.97 & 4.96/4.98/5.00 & 4.98 & 3.85/3.79/3.78 & 4.27 & 1 \\
NJU-TencentHY      & 4.76/4.97/4.97 & 4.90 & 5.00/5.00/5.00 & 5.00 & 4.14/3.71/3.68 & 4.24 & 2 \\
BJTU\_Unisound\_team & 4.69/4.80/4.80 & 4.76 & 4.72/4.71/4.84 & 4.76 & 4.02/3.85/3.77 & 4.21 & 3 \\
SenseDialog        & 3.66/3.68/3.68 & 3.67 & 4.92/4.92/4.92 & 4.92 & 4.93/3.75/3.66 & 4.06 & 4 \\
HDTLAB             & 4.34/4.01/4.60 & 4.32 & 4.24/4.41/5.00 & 4.55 & 3.74/3.37/3.48 & 3.86 & 5 \\
IUSpeech           & 2.94/2.59/2.89 & 2.81 & 2.62/2.54/3.84 & 3.00 & 2.78/3.20/3.34 & 3.07 & 6 \\
Lingcon insight    & 2.65/2.41/2.65 & 2.57 & 2.58/2.35/3.63 & 2.86 & 2.81/3.00/3.17 & 2.91 & 7 \\
Baseline           & 2.68/2.53/2.65 & 2.62 & 2.49/2.25/3.46 & 2.73 & 2.73/2.85/3.06 & 2.82 & 8 \\
\bottomrule
\end{tabular}
}
\vspace{-17pt}
\end{table*}

\begin{table}[t]
\vspace{-3pt}
\centering
\caption{Results of Track II - Full-Duplex Interaction, * indicates late submission.}
\label{tab:track2_results}
\setlength{\tabcolsep}{4pt}
\resizebox{\linewidth}{!}{
    \begin{tabular}{l c c c c c c}
    \toprule
    \multirow{2}{*}{\textbf{Team}} & \textbf{Int.} & \textbf{Rej.} & \multicolumn{2}{c}{\textbf{Delay Metric}} & \multirow{2}{*}{\textbf{Final Score}} & \multirow{2}{*}{\textbf{Rank}} \\
    \cmidrule(lr){4-5}
     & \textit{(Total)} & \textit{(Total)} & \textit{Time (s)} & \textit{Score} & & \\
    \midrule
    Cookie\_asr       & 79.3 & 72.2 & 1.260 & 79.9 & 76.6 & 1 \\
    Badcat            & 89.7 & 57.8 & 1.632 & 72.6 & 73.5 & 2 \\
    SenseDialog       & 76.4 & 60.9 & 1.237 & 80.5 & 71.0 & 3 \\
    Unity Squad*       & 68.5 & 51.2 & 1.876 & 68.6 & 61.6 & - \\
    RhythmSense       & 77.4 & 38.6 & 1.577 & 73.5 & 61.1 & 4 \\
    Lingcon Insight   & 67.6 & 38.9 & 1.127 & 83.1 & 59.2 & 5 \\
    Baseline          & 75.9 & 35.2 & 2.531 & 60.0 & 56.4 & 6 \\
    HelloWorld        & 51.3 & 36.3 & 0.624 & 100.0& 55.0 & 7 \\
    AISpeech          & 47.7 & 33.9 & 3.391 & 51.6 & 43.0 & 8 \\
    Cascade           & 28.1 & 30.9 & 1.739 & 70.7 & 37.7 & 9 \\
    \bottomrule
    \end{tabular}
}
\vspace{-15pt}
\end{table}

\section{Track I: Emotional Intelligence}
\vspace{-5pt}
This track evaluates the perception, reasoning, and generation of emotional dynamics through three tasks: 1) Emotional Trajectory Detection, which evaluates the ability to accurately identify and summarize emotional changes in multi-turn conversations; 2) Emotional Reasoning, which assesses the capacity to perceive the underlying causes of a user's emotions; and 3) Empathy Assessment, which tests the ability to generate empathetic responses in both text and audio formats.

\noindent \textbf{A. Dataset Construction.}
We employ a hybrid pipeline that combines LLM-based script generation with human performance. We utilize Gemini2.5-pro~\cite{gemini2.5} to create coherent user thought flows ($T_1{\rightarrow}T_n$) spanning 3--5 turns. Specifically, scripts for Task 1 feature dynamic trajectories in which initial emotions are disrupted by interference events, whereas Task 2 focuses on implicit causal chains in which latent triggers drive emotional state transitions. Task 3 is subsequently constructed by extracting critical emotional segments directly from these multi-turn dialogues. Finally, all data covers six balanced emotion categories and is recorded by professional actors. Detailed data statistics for both tracks are presented in Table~\ref{tab:dataset_stats}.

\vspace{2pt}
\noindent \textbf{B. Evaluation Methodology.}
We adopt an automated-human hybrid framework. \textit{Automated Evaluation:} The Qwen3-Omni-30B model ~\cite{qwen2-omni} judges Trajectory (T1), Reasoning (T2), and the Textual Empathy of T3. \textit{Human Evaluation:} A team of 20 human evaluators (divided evenly into Chinese and English groups) assesses the \textit{Emotional Appropriateness} and \textit{Audio Naturalness} of T3. All evaluators hold bachelor's degrees and possess over six months of data annotation experience. The final score is:
\begin{equation}
    \small
    \setlength{\abovedisplayskip}{3pt}
    \setlength{\belowdisplayskip}{3pt}
    \text{Score} = 0.2 S_{\text{T1}} + 0.2 S_{\text{T2}} + 0.1 S_{\text{text}} + 0.25 S_{\text{emo}} + 0.25 S_{\text{nat}}
\end{equation}
where $S_{\text{T1/T2}}$ are LLM scores for Tasks 1\&2. For Task 3, $S_{\text{text}}$ is the LLM empathy score, while $S_{\text{emo}}$ and $S_{\text{nat}}$ denote human-evaluated emotional appropriateness and naturalness.

\vspace{-5pt}
\section{Track II: Full-Duplex Interaction}
\vspace{-7pt}
This track evaluates the system's real-time decision-making capabilities during concurrent listening and speaking. Specifically, the evaluation consists of two core scenarios: 1) \textit{Interruption}, assessing responses to user interventions like follow-up questions (FQ), negation/dissatisfaction (ND), repetition requests(RR), topic switching (TS), and silence / termination (ST); and 2) \textit{Rejection}, testing robustness against non-instructional speech such as user real-time backchannels (URB), pause handling (PH), third-party speech (TPS), and speech directed at others (SDO).

\vspace{2pt}
\noindent \textbf{A. Dataset Construction.}
Similar to Track I, we employ a hybrid pipeline combining LLM-based script generation and human performance. We utilize DeepSeek~\cite{deepseek} to generate naturalistic dialogue scripts embedded with specific interaction cues (e.g., barge-ins or side-talk). Professional actors then perform these scripts to replicate authentic full-duplex dynamics. Unlike synthetic mixing, this setup ensures that interruptions occur at cognitively meaningful semantic junctures (e.g., during hesitations) rather than random timestamps, preserving natural overlap timing and prosody.

\vspace{2pt}
\noindent \textbf{B. Evaluation Methodology.}
To ensure fairness, all systems are evaluated within standardized Docker environments powered by NVIDIA RTX A6000 GPUs. The assessment covers three key dimensions: 1) Interruption, evaluating metrics such as Response Rate and Latency; 2) Rejection, evaluating metrics such as Rejection Rate and Early Interrupt Rate; and 3) Overall First Response Delay. The final score is calculated as:
\begin{equation}
    \small
    \setlength{\abovedisplayskip}{3pt}
    \setlength{\belowdisplayskip}{3pt}
    \text{Score} = 0.4 S_{\text{Int}} + 0.4 S_{\text{Rej}} + 0.2 S_{\text{Delay}}
\end{equation}
where $S_{\text{Int}}$ and $S_{\text{Rej}}$ aggregate success rates, and $S_{\text{Delay}}$ is the latency score relative to a 60-point baseline.

\vspace{-9pt}
\section{Results}
\vspace{-9pt}
The HumDial Challenge attracted over 100 registered teams, yielding 15 valid submissions.
In Track I, top teams achieved near-ceiling performance in emotional tracking and reasoning but struggled in Task 3 (Table~\ref{tab:track1_results}). This underscores that while LLMs excel at analyzing emotional logic, generating empathetic vocal and textual responses remains difficult.
In Track II, top systems exhibited diverse strengths in real-time interaction (Table~\ref{tab:track2_results}). While \textit{Badcat} achieved the highest Interruption success rate, \textit{Cookie\_asr} secured the top rank by delivering the best trade-off between low latency and robust noise rejection. However, scores for Rejection (silence maintenance) were consistently lower than Interruption, indicating that distinguishing valid user turns from background noise remains the primary hurdle for full-duplex systems. 
We plan to utilize HumDial datasets to benchmark leading commercial and open-source models, with a comprehensive comparative analysis against these submissions presented in a follow-up publication.

\bibliographystyle{IEEEbib}
\bibliography{strings,refs}

\end{document}